# Effect of Rain Scavenging on Altitudinal Distribution of Soluble Gaseous Pollutants in the Atmosphere

Tov Elperin[a*], Andrew Fominykh[a], Boris Krasovitov[a], Alexander Vikhansky[b]

[a] *Department of Mechanical Engineering, The Pearlstone Center for Aeronautical Engineering Studies, Ben-Gurion University of the Negev, P. O. B. 653, 84105, Israel*

[b] *School of Engineering and Material Science, Queen Mary, University of London, Mile End Road, London E1 4NS, UK*

## Abstract

We suggest a model of rain scavenging of soluble gaseous pollutants in the atmosphere. It is shown that below-cloud gas scavenging is determined by non-stationary convective diffusion equation with the effective Peclet number. The obtained equation was analyzed numerically in the case of log-normal droplet size distribution. Calculations of scavenging coefficient and the rates of precipitation scavenging are performed for wet removal of ammonia ($NH_3$) and sulfur dioxide ($SO_2$) from the atmosphere. It is shown that scavenging coefficient is non-stationary and height-dependent. It is found also that the scavenging coefficient strongly depends on initial concentration distribution of soluble gaseous pollutants in the atmosphere. It is shown that in the case of linear distribution of the initial concentration of gaseous pollutants whereby the initial concentration of gaseous pollutants decreases with altitude, the scavenging coefficient increases with height in the beginning of rainfall. At the later stage of the rain scavenging coefficient decreases with height in the upper below-cloud layers of the atmosphere.

## 1. Introduction

Predicting chemical composition of the atmosphere and elucidating processes which affect atmospheric chemistry is important for addressing problems related to air quality, climate and ecosystem health. Wet deposition is very important in the removal of gaseous pollutants from the atmosphere, and thus strongly affects global concentration of gaseous pollutants in the atmosphere of Earth. Atmospheric composition is controlled by natural and anthropogenic emissions of gases, their subsequent transport and removal processes. Wet deposition, including below-cloud scavenging by rains, is one of the most important removal mechanisms that control the distribution, concentration and life-time of many gaseous species in the atmosphere. Rains, through the below-cloud scavenging and aqueous-phase processes, alter the chemical composition of the atmosphere on a global scale (see, e.g. Zhang et al. 2006). Inorganic nitrogen in wet deposition is a significant source of nutrients for phytoplankton and has a direct impact on the health of estuaries and coastal water bodies (see, e.g. Mizak et al. 2005). Negative impact of $SO_2$ on visibility was indicated, e.g. by Watson (2002), Green et al. (2005) and by Tsai et al. (2007).

---

[*] Corresponding author Tel.: +972 8 6477078
Fax: +972 8 6472813
E-mail: elperin@bgu.ac.il

**Nomenclature**

| | |
|---|---|
| $a$ | raindrop radius, m |
| $c$ | total concentration of soluble trace gas in gas and liquid phases, $\text{mole} \cdot \text{m}^{-3}$ |
| $c^{(G)}$ | concentration of a soluble trace gas in a gaseous phase, $\text{mole} \cdot \text{m}^{-3}$ |
| $c_c^{(G)}$ | concentration of a soluble gas at a lower boundary of a cloud, $\text{mole} \cdot \text{m}^{-3}$ |
| $c_{gr}^{(G)}$ | concentration of a soluble gas at a ground level, $\text{mole} \cdot \text{m}^{-3}$ |
| $C^{(G)} = c^{(G)} / c_{c,0}^{(G)}$ | dimensionless concentration of trace gas in an atmosphere |
| $c^{(L)}$ | concentration of dissolved gas in a droplet, $\text{mole} \cdot \text{m}^{-3}$ |
| $d$ | raindrop diameter, m |
| $D_G$ | coefficient of diffusion in a gaseous phase, $\text{m}^2 \cdot \text{s}^{-1}$ |
| $D$ | effective coefficient of diffusion, $\text{m}^2 \cdot \text{s}^{-1}$ |
| $m$ | parameter of solubility |
| $R$ | rainfall rate, $\text{m} \cdot \text{s}^{-1}$ |
| $L$ | distance between ground and lower boundary of a cloud, m |
| $q_c$ | flux of dissolved gas, transferred by rain droplets, $\text{mole} \cdot \text{m}^{-2} \cdot \text{s}^{-1}$ |
| $\text{Pe} = UL / D$ | Peclet number |
| $\text{Re} = u \cdot d / \nu_G$ | external flow Reynolds number for a moving droplet |
| $\text{Sc} = \nu_G / D_G$ | Schmidt number |
| $\text{Sh} = \beta \cdot d / D_G$ | Sherwood number |
| $t$ | time, s |
| $T = tU / L$ | dimensionless time |
| $u$ | velocity of a droplet, $\text{m} \cdot \text{s}^{-1}$ |
| $U$ | "wash-down" front velocity, $\text{m} \cdot \text{s}^{-1}$ |
| $z$ | coordinate in a vertical direction, m |

*Greek symbols*

| | |
|---|---|
| $\beta$ | coefficient of mass transfer $\text{m} \cdot \text{s}^{-1}$ |
| $\phi$ | volume fraction of droplets in the air |
| $\mu$ | dynamic viscosity of a fluid, $\text{kg} \cdot \text{m}^{-1} \cdot \text{s}^{-1}$ |
| $\tau_{ch}$ | characteristic time of concentration change in a gaseous phase, s, |
| $\tau_D$ | characteristic time of diffusion process, s, |
| $\Lambda$ | scavenging coefficient, $\text{s}^{-1}$ |

*Subscripts*

| | |
|---|---|
| 0 | initial value |
| $c$ | value at a lower boundary of a cloud |
| $gr$ | value at a ground |
| $G$ | gaseous phase |
| $L$ | liquid phase |

Mixed with water or reacting with other chemicals in the air $SO_2$ has negative health effect. Gas scavenging by rain includes absorption of $SO_2$, $NH_3$ and other gases. Sulfur dioxide $SO_2$ is emitted from smokestacks as a result of various fossil fuels combustion, e.g., crude oil and coal. The main source of atmospheric ammonia is agriculture (see, e.g. Van Der Hoek, 1998), and the remaining minor sources are industries, humans, pets, wild animals, landfills and households products. The contribution of vehicles to non-agricultural $NH_3$ emissions has been considered to be negligible until 1995. In the last years, however, an increase of $NH_3$ emission due to introduction of petrol-engine vehicles equipped with catalytic converters has been reported in the literature (see Fraser and Cass, 1998). Concentration measurements of $SO_2$, $NH_3$ and other trace gases in the atmospheric boundary layer revealed vertical (altitudinal) dependence of the concentrations (see Georgii 1978; Gravenhorst et al. 1978; Georgii and Müller 1974). Concentration of gases which are not associated with photosynthesis, e.g. $SO_2$ and $NH_3$, has a maximum at the Earth surface and decreases with height over the continents. The concentration of $NH_3$ over the continents decreases rapidly with altitude, reaching a constant background concentration at the altitudes of about 1500 m above the ground in winter and at the altitudes of about 3000 m above the ground on warm days (see Georgii and Müller 1974; Georgii 1978). On warm days the ground concentration of $NH_3$ is considerably higher than that on the cold days. Sulfur dioxide concentration in the ABL (atmospheric boundary layer) is higher during winter than during summer because of the higher anthropogenic $SO_2$ production. In contrast to the concentrations of $SO_2$ and $NH_3$ over the continents, the profiles of concentration of these gases over the ocean have minimum at the ocean surface. This phenomenon is explained by a high solubility of $SO_2$ and $NH_3$ in a sea water whereby the ocean acts as a sink of soluble gases (see Georgii and Müller 1974; Georgii 1978). Information about the evolution of the vertical profile of soluble gases with time allows calculating fluxes of these gases in an the ABL. Vertical transport of soluble gases in the ABL is an integral part of the atmospheric transport of gases and is important for understanding the global distribution pattern of soluble trace gases. An improved understanding of the cycle of soluble gases is also essential for the analysis of global climate change. Clouds and rains play essential role in vertical redistribution of $SO_2$,

$NH_3$ and other soluble gases in the atmosphere. Scavenging of soluble gases, e.g., $SO_2$, $NH_3$ by rains contributes to the evolution of vertical distribution of these gases. At the same time the existence of vertical gradients of the soluble gases in the atmosphere affects the rate of gas absorption by rain droplets (see Elperin et al. 2009). Note that the existing models of global transport in the atmosphere do not take into account the influence of rains on biogeochemical cycles of different gases.

In spite of a large number of publications devoted to soluble gases scavenging by clouds (see, e.g., Elperin et al. 2007 and Elperin et al. 2008 and references therein) there are only a few studies on scavenging of these gases by rains. Hales (1972, 2002), Hales et al. (1973) and Slinn (1974) considered removal of soluble pollutant gases from gas plumes. Hales (1972, 2002) and Hales et al. (1973) assumed that concentration of the dissolved gas in a droplet is equal to concentration of saturation in liquid, corresponding to concentration of a trace soluble gas in the atmosphere at a certain height. Hales (1972) showed that if a drop falls through a plume and emerges into a clean air before reaching the ground, it may release most of the soluble gaseous pollutants that has been removed from more polluted regions. The significance of this effect is lowering the altitude of the regions with increased concentration of soluble gaseous pollutants under the influence of rain. Hales (2002) considered a set of equations that correspond to five kinds of Gaussian plume formulation. This approach is valid for low gradients of concentration in a gaseous phase and for absorption of gases with low solubility, when

$$\tau << \tau_{ch} \qquad (1)$$

where $\tau = a \cdot m / \beta$ is a characteristic diffusion time and $\tau_{ch} = c_{gr}^{(G)} / \left( u \cdot \left| dc^{(G)} / dz \right| \right)$ is a characteristic time of concentration change in a gaseous phase. Slinn (1974) showed that plume's "washdown" velocity can be calculated as $w = I_0 \cdot H'$, where $I_0$ is a rainfall rate and $H'$ is a dimensionless Henry's constant. The latter approach is valid for uniform droplet size distribution in the rain. Zhang et al. (2006) investigated numerically gas scavenging by drizzle developed from low-level, warm stratiform clouds using the approach of Ackerman et al (1995), developed for modeling condensation nuclei and water droplets size distributions, and considered precipitation rates in the range from 0.01 $\text{mm} \cdot \text{h}^{-1}$ up to 0.06 $\text{mm} \cdot \text{h}^{-1}$. The conclusion of this study was that total droplet surface area is more appropriate than the precipitation rate for parameterizing scavenging coefficients, especially when precipitation has a large fraction of small drops. Different aspects of soluble gaseous pollutants scavenging by rain droplets were discussed by Pruppacher and Klett (1997), Wurzler (1998), Stefan and Mircea (2003), Slinn (1977), Calderon et al. (2008), Asman (1995), Mircea et al. (2000, 2004), Kumar (1985), Levine and Schwartz (1982), Dana et al. (1975), Elperin and Fominykh (2005). Asman (1995) investigated absorption of highly soluble gases by rain using the approximation of infinite solubility of absorbate in the absorbent and assuming that distribution of soluble gas in the atmosphere during the rain is time-dependent and uniform. The latter assumption allowed calculating numerically the dependence of the scavenging coefficient on the rainfall rate in the atmosphere. Power law dependence of the scavenging coefficient on the rainfall rate for ammonia absorption by rain, which was predicted by Asman (1995) theoretically, was confirmed experimentally by Mizak et al. (2005). All the above studies did not account for the dependence of scavenging coefficient on height, time and initial profile of soluble gas in the atmosphere. In this study we investigate the influence of the altitude absorbate inhomogenity in a gaseous phase on the rate of soluble gas scavenging by falling rain droplets. The problem is reduced to the equation of non-stationary convective diffusion with the effective Peclet number that depends on droplets size distribution (DSD). The obtained equation was solved numerically for log-normal DSD with Feingold-Levin parameterization (Feingold and Levin, 1986), and time and altitude dependence of the scavenging coefficient was analyzed.

## 2. Description of the model

In this study we consider absorption of a moderately soluble gas from a mixture containing inert gas by falling rain droplets. At time t = 0 rain droplets begin to fall and absorb gaseous pollutants (trace gases) from the atmosphere. It is assumed that the initial concentration of the dissolved trace gas in rain droplets is equal to the concentration of saturation in liquid corresponding to concentration of a trace soluble gas in a cloud and that the initial distribution (at time t = 0) of soluble trace gas in the atmosphere is known.

It must be noted that for moderately soluble gases, only a small fraction of the gas dissolves in the cloud water. Therefore the concentration of the moderately soluble gas in the interstitial air in a cloud is close to the concentration of the soluble gas in the below-cloud atmosphere immediately adjacent to the cloud. Since the residence time of droplets in the cloud is large, the equilibrium is established between the concentration of a moderately soluble gas in the interstitial air and concentration of the dissolved gas in the cloud droplets (see Asman 1995).

The goal of this study is to determine an evolution of concentration distribution of soluble trace gases in the atmosphere below the cloud under the influence of gas scavenging by falling rain droplets. Our analysis is not restricted to gases with low solubility and is valid for all gases which obey Henri's law. The suggested model is not constrained by a magnitude of a gradient of the soluble trace gas concentration in a gaseous phase. Following the approach suggested by Hales (1972, 2002) and Hales et al. (1973), time derivative of the mixed-average concentration of the dissolved gas in a falling droplet can be written as follows:

$$\frac{dc^{(L)}}{dt} = \frac{1}{\tau_D}\left(mc^{(G)} - c^{(L)}\right), \qquad (2)$$

where $\tau_D = (a \cdot m)/(3\beta)$ is a characteristic diffusion time, m is a solubility parameter, $\beta$ - mass transfer coefficient in a gaseous phase, $c^{(L)}$ - mixed-average concentration of the dissolved gas in a droplet, [mole·l$^{-1}$], $c^{(G)}$ - concentration of a soluble gaseous pollutant in a gaseous phase, [mole·l$^{-1}$], a - raindrop radius. Dimensionless mass transfer coefficient for a falling droplet in a case of gaseous phase controlled mass transfer, $Sh = \beta \cdot d / D_G$, can be written as follows (see, e.g. Seinfeld (2006)):

$$Sh = 2 + 0.6 \cdot Re^{1/2} \cdot Sc^{1/3}, \qquad (3)$$

where $Re = u \cdot d / \nu_G$, $Sc = \nu_G / D_G$. For small $\tau_D$ Eq. (2) yields:

$$c^{(L)} = m\left(c^{(G)} - \tau_D \frac{dc^{(G)}}{dt}\right). \qquad (4)$$

Total concentration of soluble gaseous pollutant in gaseous and liquid phases reads:

$$c = \left(1-\phi\right)c^{(G)} + \phi c^{(L)} \qquad (5)$$

As can be seen from the Eq. (4) in the case when $\frac{\tau_D}{c^{(G)}}\left|\frac{dc^{(G)}}{dt}\right| << 1$ Eqs (4)-(5) yield:

$$c = c^{(G)}\left[(1-\phi) + m\phi\right], \qquad (6)$$

where $\phi$ - volume fraction of droplets in the air. The total flux of the dissolved gas transferred by rain droplets is determined by the following expression:

$$q_c = \phi \cdot u \cdot c^{(L)}, \qquad (7)$$

where $u$ - velocity of a droplet, $c^{(L)}$ - concentration of dissolved gas in a droplet. Using Eqs. (4) and (7) we obtain:

$$q_c = m\phi u\left(c^{(G)} - \tau_D \frac{dc^{(G)}}{dt}\right). \qquad (8)$$

Equation of mass balance for soluble trace gas in the gaseous and liquid phases is as follows:

$$\frac{\partial c}{\partial t} = -\frac{\partial q_c}{\partial z}. \qquad (9)$$

Combining Eqs. (4) – (9) we obtain the following convective diffusion equation:

$$\frac{\partial c^{(G)}}{\partial t} + U\frac{\partial c^{(G)}}{\partial z} = D\frac{\partial^2 c^{(G)}}{\partial z^2}, \qquad (10)$$

where $\quad U = \dfrac{m\phi u}{(1-\phi) + m\phi}, \qquad D = \dfrac{m\phi\tau \cdot u'^2}{(1-\phi) + m\phi},$ $u' = u - U$ and $u >> U$. The term in the right-hand side of Eq. (10) arises because we do not make a simplifying assumption about equality between the instantaneous concentration of the dissolved gas in a droplet and concentration of saturation in liquid corresponding to the concentration of a trace soluble gas in an atmosphere at a given height. In other words, Eq. (10) is valid when a characteristic diffusion time and a characteristic time of concentration change in a gaseous are of the same order of magnitude. For example, for $SO_2$ absorption by 1.2 mm diameter water droplet, $U = 1.2\times10^{-4}\,\text{m}\cdot\text{s}^{-1}$, $\tau_D = 4.52\times10^{-2}\,\text{sec}$, $D = 2.17\times10^{-5}\,\text{m}^2\cdot\text{s}^{-1}$, and for ammonia absorption by 1.2 mm diameter water droplet $U = 6.06 \times 10^{-3}\,\mathrm{m \cdot s^{-1}}$, $\tau_D = 2.463\,\mathrm{sec}$, $D = 5.97 \times 10^{-2}\,\mathrm{m^2 \cdot s^{-1}}$. In the present model it was assumed that the soluble gaseous species are molecularly dissolved in water droplets, and the molecules of these species do not dissociates into ions in the liquid phase (see Seinfeld and Pandis 2006, Chapter 7). The initial and boundary conditions for Eq. (10) are as follows:

$$t = 0, \quad c^{(G)} = f(z) \tag{11}$$

$$z = 0, \quad c^{(G)} = c_{c,0}^{(G)}, \tag{12}$$

$$z = L, \quad \frac{\partial c^{(G)}}{\partial z} = 0, \tag{13}$$

where $L$ - distance between the ground and the lower boundary of a cloud. Equation of non-stationary convective diffusion (10) with initial and boundary conditions (11) - (13) (see, e.g. Leij and Toride 1998) describes evolution of solvable trace gas distribution in the atmosphere under the influence of rain. Equation (13) is a condition of ground impermeability for soluble gases. The volume fraction of a liquid phase $\phi$ in Eq. (10) determines the intensity of rain, $R = u\phi$ and $U$ is the "wash-down" front velocity. Equation (10) implies that trace gas in the atmosphere is scavenged with a "wash-down" velocity $U$ and is smeared by diffusion. Equations (10) - (13) can be rewritten in the following form:

$$\frac{\partial C^{(G)}}{\partial T} + \frac{\partial C^{(G)}}{\partial \eta} = \frac{1}{\mathrm{Pe}} \cdot \frac{\partial^2 C^{(G)}}{\partial \eta^2} \tag{14}$$

$$T = 0, \quad C^{(G)} = f(\eta), \tag{15}$$

$$\eta = 0, \quad C^{(G)} = 1, \tag{16}$$

$$\eta = 1, \quad \frac{\partial C^{(G)}}{\partial \eta} = 0, \tag{17}$$

Where $\mathrm{Pe} = UL/D$, $T = tU/L$, $C^{(G)} = c^{(G)}/c_{c,0}^{(G)}$.
Assuming that the dependence of the terminal fall velocity of a liquid droplet depends on its diameter is as follows (see Kessler, 1969):

$$u = c_1 \cdot d^{1/2}, \tag{18}$$

where $c_1 = 130\,[\mathrm{m^{\frac{1}{2}} \cdot s^{-1}}]$, and u and d are expressed in SI units, we obtain the following formula for Peclet number:

$$\mathrm{Pe} = \frac{U \cdot L}{D} = \frac{6D_G \cdot L}{m \cdot c_1 \cdot d^{5/2}} \cdot \left[ 2 + 0.6 \cdot \frac{c_1^{1/2} \cdot d^{3/4}}{\nu_G^{1/6} \cdot D_G^{1/3}} \right] \tag{19}$$

For ammonia $D_{G,\mathrm{NH_3}} = 0.21 \cdot 10^{-4}\ \mathrm{m^2 \cdot s^{-1}}$, $m_{\mathrm{NH_3}} = 1515$, $\nu_{G,air} = 15 \cdot 10^{-6}\ \mathrm{m^2 \cdot s^{-1}}$. Consequently, for ammonia scavenging by 1 mm diameter water droplets falling in the air and for L = 1000m, Pe = 217. In a case of sulfur dioxide $D_{G,\mathrm{SO_2}} = 0.122 \cdot 10^{-4}\ \mathrm{m^2 \cdot s^{-1}}$, $m_{\mathrm{SO_2}} = 30$ and $Pe = 5.53 \cdot 10^3$. If the initial distribution of a trace gas in the atmosphere is linear, the boundary condition given by Eq. (15) becomes

$$T = 0, \quad C^{(G)} = 1 + \left( c_{gr,0}^{(G)}/c_{c,0}^{(G)} - 1 \right) \cdot \eta. \tag{20}$$

## 3. Results and discussions

The above model of atmospheric trace gases scavenging by liquid precipitation was applied to study the evolution of trace soluble gas concentration in the atmosphere caused by rain. Results of numerical solution of Eqs. (14) – (17) with linear initial distribution of soluble trace gas in the atmosphere for $c_{gr,0}^{(G)}/c_{c,0}^{(G)} = 2$ are presented at Figs. 1 and 2.

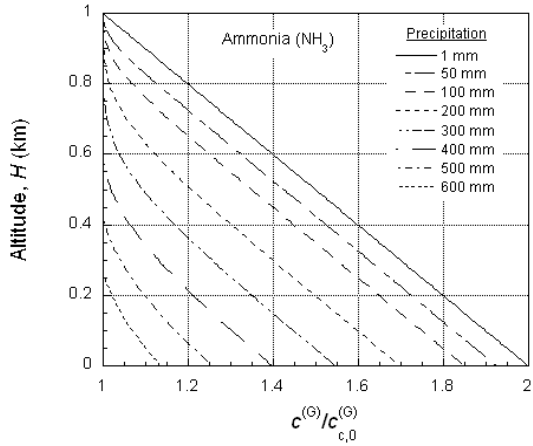

**Fig. 1.** Evolution of ammonia distribution in the atmosphere caused by rain scavenging

Calculations are performed for scavenging of $NH_3$ and $SO_2$ by rain. Wet removal of soluble gases from the atmosphere strongly depends on the raindrops diameter that is determined by droplet size distribution (DSD). In our calculations we assumed the log-normal size distribution of raindrops with Feingold and Levin parameterization (Feingold and Levin, 1986) based on the long-time measurements of rain drops size spectra in Israel:

$$n(d) = \frac{N_d}{\sqrt{2\pi} d \ln \sigma} \exp\left[ -\frac{(\ln d - \ln d_r)^2}{2(\ln \sigma)^2} \right],$$

$$(21)$$

$$N_d = 172 R^{0.22} \left( m^{-3} \right); \qquad d_r = 0.72 R^{0.23} \left( mm \right);$$

$$\sigma = 1.43 - 3 \cdot 10^{-4} R,$$

where $R$ is the rain intensity (mm $h^{-1}$). Volumetric fraction of raindrops in the atmosphere was assumed to be equal to $10^{-6}$.

Inspection of Figs. 1 and 2 shows that the larger is the solubility of the trace gas in water, the smaller it is the quantity of precipitation required to washout it. For instance, inspection of Fig. 1 reveals that approximately 600 mm of precipitation can wash out 1 km of atmosphere from the ammonia gas.

At the same time for wet removal of sulfur dioxide from the atmosphere of the same altitude the considerably higher amount of precipitation is required.

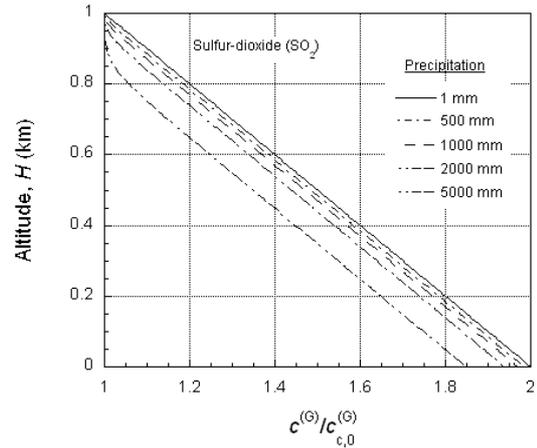

**Fig. 2.** Evolution of sulfur dioxide distribution in the atmosphere caused by rain scavenging

In the calculations the initial concentration of dissolved trace gas in rain drops is assumed equal to the concentration of saturation in a liquid corresponding to the concentration of a trace soluble gas in a cloud. Therefore the soluble gas in the below-cloud atmosphere can be washed down only up to the concentration of soluble gas in the interstitial air in a cloud. Note that for gaseous pollutants their concentration at the ground is always larger than the concentration in a cloud (see e.g., Georgii and Müller, 1974; Georgii, 1978; Gravenhorst et al., 1978). Inspection of Figs. 1–2 shows that the thickness of the layer "washed down" by precipitation strongly depends on the rainfall amount and also depends on the gas solubility.

Using the obtained numerical solution of the equation (14) with the boundary conditions (15)–(17) we also calculated the scavenging coefficient for soluble trace gas absorption from the atmosphere:

$$\Lambda = -\frac{1}{c^{(G)}} \frac{\partial c^{(G)}}{\partial t} . \qquad (22)$$

The dependence of the scavenging coefficient vs. altitude in the case of ammonia wash out is shown in Figs. 3 – 4.

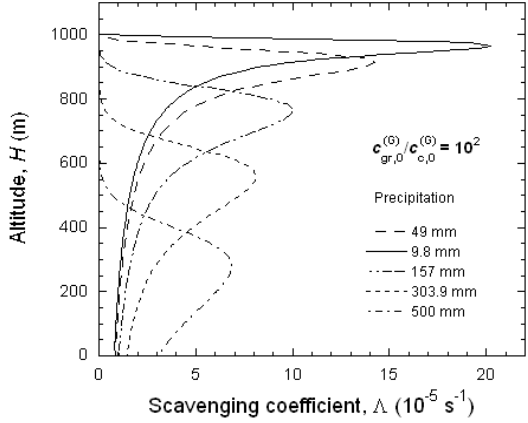

**Fig. 3.** Dependence of scavenging coefficient vs. altitude for ammonia wash out (linear initial distribution of ammonia in the atmosphere $c_{gr,0}^{(G)} \big/ c_{c,0}^{(G)} = 100.0$ ).

The previous studies have showed that the scavenging coefficient which is measured or calculated under the assumption of the uniform soluble trace gas distribution may not accurately predict wet deposition of soluble trace gases in the presence of a gradient of concentration of trace gases in the atmosphere (see e.g. Assman, 1995; Mizak et al., 2005; Calderon et al., 2008).

The suggested model takes into account the initial concentration gradient of soluble species in the atmosphere.

Calculations were performed for linear initial distribution of ammonia in the atmosphere and different ratios of the ammonia concentration in a cloud and at the ground: $c_{gr,0}^{(G)} \big/ c_{c,0}^{(G)} = 100.0$ (see Fig. 3) and $c_{gr,0}^{(G)} \big/ c_{c,0}^{(G)} = 2.0$ (see Fig. 4).

Inspection of Figs. 3 – 4 shows that scavenging coefficient strongly depends on the initial distribution of soluble trace gas concentration in the atmosphere.

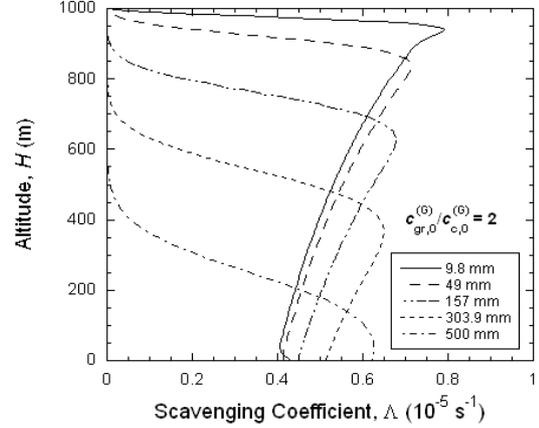

**Fig. 4.** Dependence of scavenging coefficient vs. altitude for ammonia wash out (linear initial distribution of ammonia in the atmosphere $c_{gr,0}^{(G)} \big/ c_{c,0}^{(G)} = 2.0$ )

As can be seen from these plots the scavenging coefficient increases with the increase of the soluble species concentration gradient. The analysis of the plots on Figs. 3 – 4 shows that the high values of the scavenging coefficient $\Lambda$ in the below-cloud atmosphere immediately adjacent to the cloud at the early stage of a rain is explained by high rates of gas absorption by falling rain droplets. High rates of mass transfer between the rain droplets and soluble gas are caused by thin concentration boundary layers in droplets and in a gaseous phase at the initial stage of gas-liquid contact. For linear profile of soluble gas in the atmosphere, at the early stage of a rain, the scavenging coefficient increases with height. In the case when $\tau_D \ll \tau_{ch}$ the mass flux to the rain droplets falling in an atmosphere with a non-uniform initial concentration profile of soluble gas is constant. Therefore the rate of change of concentration in a gaseous phase $\partial c^{(G)} / \partial t$ is constant. At the same time the distribution of the concentration in an atmosphere $c^{(g)}$ decreases with height. Consequently $\Lambda$ increases with height at the early stage of rain whereby the initial concentration profile of the soluble gas in the atmosphere is not disturbed significantly. Scavenging of soluble gas begins in the upper atmosphere and the front of scavenging propagates downwards with the "wash down" velocity that is proportional to Henry's constant and rain intensity (see Eq. 10). Concentration of a soluble gas in the below-cloud layer decreases to the concentration of a soluble gas in the interstitial air in a cloud. The subsequent rain droplets fall in the below-cloud atmosphere without absorbing soluble gas. This explains the decrease of the scavenging coefficient in the upper below-cloud layers of the atmosphere at the later stages of rain whereby the initial concentration profile of the soluble gas in the atmosphere changes significantly. Note that the soluble gas in the below-cloud layer is washed out only to the concentration of the soluble gas in the interstitial air in a cloud. At the ground the value of the scavenging coefficient increases with time because the concentration at the ground decreases faster than the rate of concentration change.

Dependences of scavenging coefficient on the rate of precipitation are shown in Figs. 5 and 6. The dependences of the scavenging coefficient on rain intensity are plotted for the early stage of rain (Fig. 5) as well as for the advanced stage of rain (Fig. 6).

As can be seen from these plots the scavenging coefficient increases with rain intensity increase. In spite of the numerous theoretical calculations and measurements of scavenging coefficient available in the literature (see e.g., Beilke, 1970; Sperber and Hameed, 1986; Shimshock and De Pena, 1989;

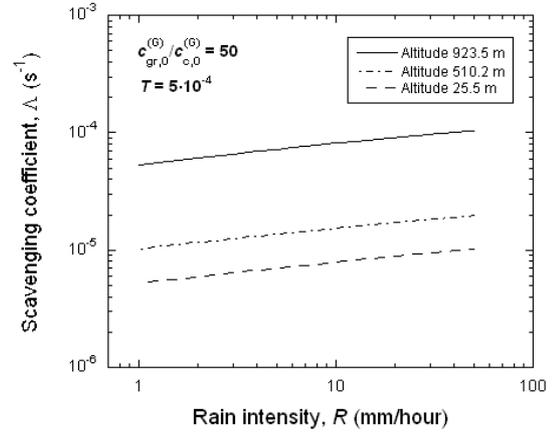

**Fig. 5.** Dependence of scavenging coefficient vs. rain intensity for ammonia wash out at the early stage of rain (dimensionless time $T = 5 \cdot 10^{-4}$; linear initial distribution of ammonia in the atmosphere $c_{gr,0}^{(g)} / c_{c,0}^{(g)} = 50.0$).

Renard et al., 2004; Mizak et al., 2005) the comparison of the predicted values of scavenging coefficient with those calculated from the measured concentrations of ammonia in rainwater reveals large discrepancies.

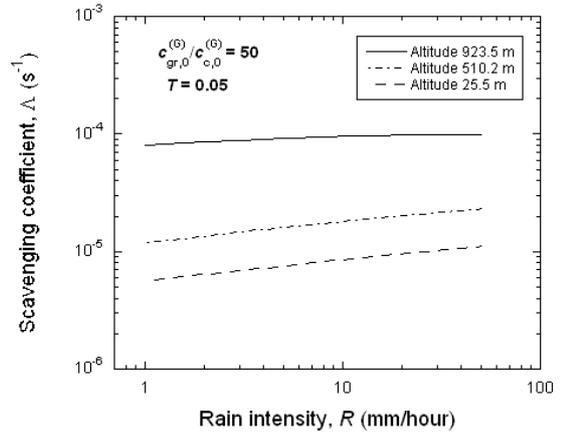

**Fig. 6.** Dependence of scavenging coefficient vs. rain intensity for ammonia wash out at the later stage of rain (dimensionless time $T = 0.05$; linear initial distribution of ammonia in the atmosphere $c_{gr,0}^{(g)} / c_{c,0}^{(g)} = 50.0$).

In particular, for wet deposition of ammonia very different values of scavenging coefficient are reported in the literature, in the range from less that $10^{-5}$ $(s^{-1})$ (Sperber and Hameed, 1986) to larger than $10^{-3}$ $(s^{-1})$ (in Mizak et al., 2005) are reported. This large scatter of data is mentioned in several studies and reviews (see e.g., Renard et al., 2004).

The wide range of variation of the magnitude of scavenging coefficient is caused by dependence of scavenging coefficient on the altitude, time, initial gradient of the soluble gas concentration in the below-cloud atmosphere, droplet size distribution as well as on meteorological conditions (wind, temperature etc.) and difficulties associated with evaluating scavenging coefficient from the experiments.

## Conclusions

In this study we developed a model for scavenging of soluble trace gases in the atmosphere by rain. It is shown that gas scavenging is determined by non-stationary convective diffusion equation with the effective Peclet number that depends on droplet size distribution (DSD). The obtained equation was analyzed numerically in the case of log-normal DSD with Feingold-Levin parameterization (Feingold and Levin, 1986). The simple form of the obtained equation allows analyzing the dependence of the rate of soluble gas scavenging on different parameters, e.g. rain intensity, gas solubility, gradient of absorbate concentration in a gaseous phase etc. Using the developed model we calculated scavenging coefficient and the rates of scavenging of different trace gases ($SO_2$ and $NH_3$). The obtained results can be summarized as follows:

1. It is demonstrated that scavenging coefficient for the wash out of soluble atmospheric gases by rain is time-dependent. It is shown that value of scavenging coefficient at the ground increases with time whereas the value of scavenging coefficient in the below-cloud atmosphere immediately adjacent to the cloud decreases with the amount of precipitation.

2. It is shown that scavenging coefficient in the atmosphere is height-dependent. Scavenging of soluble gas begins in the upper atmosphere and scavenging front propagates downwards with "wash down" velocity and is smeared by diffusion. We have found that in the case of linear initial distribution of concentration of gaseous pollutants whereby the initial concentration of gaseous pollutants decreases with altitude, the scavenging coefficient $\Lambda_c$ increases with height at an early stage of rain. At the advanced stage of rain scavenging coefficient decreases with height in the upper below-cloud layers of the atmosphere.

3. It is found that scavenging coefficient strongly depends on the initial distribution of soluble trace gas concentration in the atmosphere. Calculations performed for linear distribution of the soluble gaseous species in the atmosphere show that the scavenging coefficient increases with the increase of soluble species gradient.

The developed model can be used for the analysis of precipitation scavenging of hazardous gases in the atmosphere by rain and for validating advanced models for predicting scavenging of soluble gases by rain.